# Counterions and water molecules in charged silicon nanochannels: the influence of surface charge discreteness

Yinghua Qiu and Yunfei Chen*

*School of Mechanical Engineering and Jiangsu Key Laboratory for Design and Manufacture of Micro-Nano Biomedical Instruments, Southeast University, Nanjing, 211189,China*

yinghuaqiu@seu.edu.cn

Corresponding author: yunfeichen@seu.edu.cn, TEL:086-025-52090518(O)

# Counterions and water molecules in charged silicon nanochannels: the influence of surface charge discreteness

In order to detect the effect of the surface charge discreteness on the properties at the solid-liquid interface, molecular dynamics simulation model taking consideration of the vibration of wall atoms was used to investigate the ion and water performance under different charge distributions. Through the comparison between simulation results and the theoretical prediction, it was found that, with the degree of discreteness increasing, much more counterions were attracted to the surface. These ions formed a denser accumulating layer which located much nearer to the surface and caused charge inversion. The ions in this layer were non-hydrated or partially hydrated. When a voltage was applied across the nanochannel, this dense accumulating layer did not move unlike the ions near uniformly charged surface. From the water density profiles obtained in nanochannels with different surface charge distributions, the influence of the surface charge discreteness on the water distributions could be neglected.

**Introduction:**

At solid-liquid interface, the solid surface will be charged due to the adsorption of ions from solution or dissociation of surface groups[1]. For example: in aqueous solution, silicon surfaces are usually negatively charged due to the dissociation of protons from the hydroxyl groups on them[2]. The amount and distribution of the surface charges have close relationship with the environment at the interface[3]. The surface charges developed at the solid-liquid interface are elementary charges in real world. They show discreteness on the solid surface, which is different from the mean field theory[1]. Due to the tiny scale, the direct observation of the distribution of surface charges is hard to realize. However, because the electrostatic interaction is a strong long-rang interaction, it has great influence on the ions and water distributions at the interface and further to lots of domains, such as the stability of colloids [4], electro-osmotic flow, and electrophoresis[5].

As the size of interface down to nanoscale, the effect of interface and size become more apparent. The surface charges affect the property of fluid confined in nanochannel larger. In experiment aspect, Loh et al. [6] found that the transverse distribution of ions on the mica surface were mainly above the $Al^{3+}$ ions in mica using FM-AFM. Stein et al.[7] discovered the flow of aqueous electrolyte with low concentration confined in nanochannels were surface charge governed. In simulation aspect, Gong et al. [8] realized the selective transportation of ions through carbon nanotube by controlling the location of charge in the model. Shen et al. [9] studied the conformation of biomolecules in electrolyte and found the structure of DNA molecules were changed by the selectively adsorption of $Na^+$ and $Rb^+$ ions on the molecule surface. So, it is of great importance to conduct the research on the water and ion distribution in nanochannels in order to improve the nanofluidic technology and understanding of interface.

Silicon that exists extensively on earth is widely used in the nano-devices[10]. With the fast development of the fabrication technology, lab-on-a-chip technology has attracted lots of attention, due to its potential application in rapid single molecule sensing[11, 12] and material separation[13]. While, the control of electro-osmotic flow in nano/micro channels are the key to the lab-on-a-chip technology which character directly influenced by the water and ion distribution near the solid surface. Rong et al. [14] reported the electrostatic gating of proton transport within aligned mesoporous silica thin film and observed that surface-charge-mediated transport is dominant at low proton concentrations. Karnik et al. [15] designed a field-effect-control nanofluidic transistor circuit to sense the protein molecules in which the surface potential has large effect on the ion distribution in the nanochannel. He *et al.* [16] measured the ionic current in charged conical nanopores and found that the surface charges can cause current rectification when the direction of the electric field in the pore reverses. So, the research on the ion and water distribution can also provide advice to the design of lab-on-a-chip.

In theoretical aspect, Poisson-Boltzmann (PB) equation is a mean field theory to describe the ion distribution at the solid-liquid interface[1]. Due to the simple form and convenient application, PB equation is widely used to solve the problem about dilute solution with monovalent ions in contact to weakly charged surfaces[17]. However, because the correlation among ions and surface charges[18], ion size[19], the non-uniform dielectric constant[20] and the surface charge discreteness[21] are neglected in it, this theory fails down when the solution concentration is high or the surface is strongly charged or multivalent counterions appear at the interface. Lots of counterintuitive phenomena have been investigated, such as charge inversion[18] and like-charge attraction[22], which cannot be explained by PB theory.

The appearance of charge inversion and like-charge attraction is mainly attributed to the correlation among ions and surface charges. The strong correlated theory raised by Shkovskii[18] gives some description. In addition, the size of ions[23] and solvent molecules[24] and the surface charge discreteness[25] are found to enhance this phenomenon. Although these are counterintuitive, they may be used in bioengineering, such as DNA therapy. Due to the negatively charged cell membrane and DNA molecules, if charge inversion developed, DNA molecules can easily translocate into the inner of the cell[26, 27].

Molecular dynamics simulation offers an effective tool to discover the mystery of EDL at interface. In this work, we conducted two MD simulations with different surface charge distributions and the numerical solution of PB equation to explore the effect of the surface charge discreteness on the liquid properties at interface.

**MD details:**

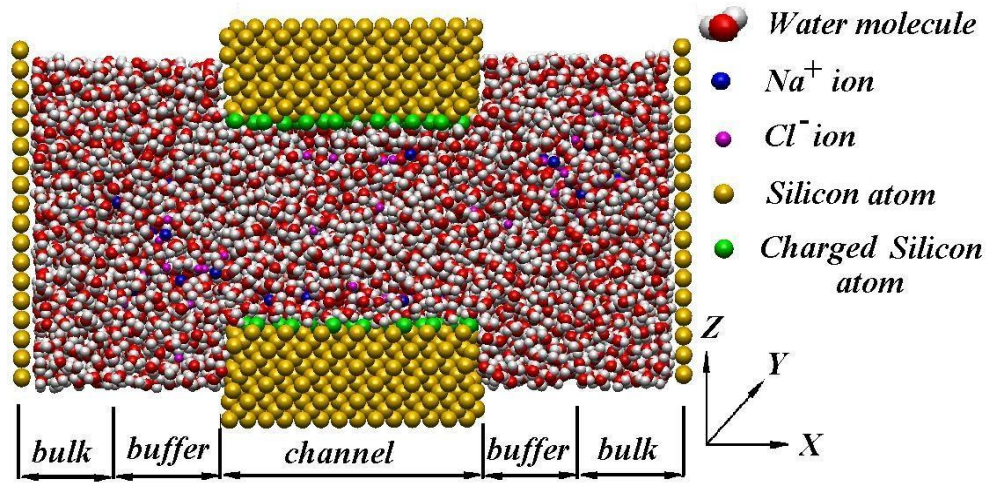

Figure 1. Schematic diagram of the MD model.

The schematic diagram of the model[28] used in the simulations is exhibited in Fig. 1, which is assumed to be infinite in the *y* direction using periodic boundary conditions. The middle part of the model is the nanochannel with 3.2 nm in height that we want to simulate and where we get the final results. Both the upper and the bottom solid walls are composed of twelve-layer silicon atoms oriented in the (100) direction, of which eight-layer silicon atoms nearest to the solution are allowed to thermally vibrate near their equilibrium positions, while the outer four layer silicon atoms are frozen without thermal vibration. There are a bulk and buffer regions on each side of the channel. Outside the bulk regions, there are a layer of sparse silicon atoms constructed as a solid boundary to the aqueous solution. The bulk region is filled with a predefined salt solution to provide enough ions and water molecules. The buffer region is used to judge whether the system achieves equilibrium. In our simulation, once the salt concentration in the buffer region is not changed with the simulation time and keeps at a predefined concentration for enough long time, the system is believed to reach equilibrium. The following simulations are used to get the final results.

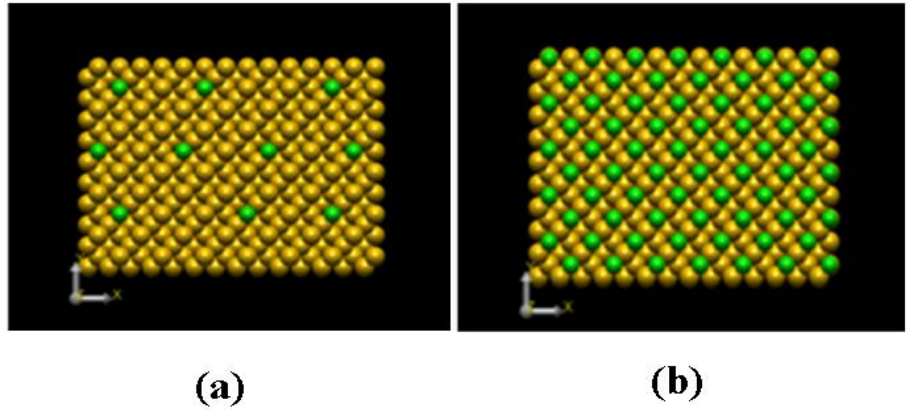

(a)   (b)

Figure 2. The surface charge distributions used in MD simulations, (a) discrete charge distribution, (b) uiniform charge distribution.

In the simulations, two cases were used to consider the surface charge discreteness i.e. discrete case and uniform case. The surface charge density was set as $-0.15$ C/m$^2$, but with two different distributions as shown in Fig. 2. The surfaces in discrete case are discretely charged with one elementary charge on one charged atom. In uniform case, the surface is uniformly charged by artificially distributing a certain number of charges to the silicon atoms on the channel inner surfaces, and each surface atom has 1/7 elementary charge.

At the beginning of each simulation, pure water was filled in the buffer and channel regions, and standard NaCl solution with 1M was filled in the two bulk regions on the two ends of the model. The standard solution was produced by another program previously in order to keep it at an equilibrium state with a given temperature. Once the simulation started, ions would diffuse from the bulk regions to the nanochannel until the concentration gradients disappeared. The concentration gradients could be evaluated from the buffer regions. Before the system reached equilibrium, the solution in the two bulk regions was replaced periodically by the standard solution. With two bulk regions, more ions could be supplied and the system can reach equilibrium much faster.

In our simulations, the TIP4P model was selected to simulate the water molecules and the SETTLE algorithm [29] was chosen to maintain the water geometry. The Lennard-Jones (LJ) potential[30] was used to describe the interactions between different atoms, except hydrogen-X pairs (X is an atom species in the solution) and silicon-silicon pairs. The Stillinger-Weber (SW) potential [31] was used to describe the interaction among the silicon atoms. The electrostatic interactions among ions, water molecules and surfaces charges were modeled by Ewald summation algorithm[32]. The motion equations were integrated by the leap-frog algorithm with time step of 2.0 fs. The solution system was maintained at 298 K by Berendesen thermostat [33] and the silicon walls were maintained at that constant temperature using a damped method [34]. The first run lasting 6 ns was used to equilibrate the system. Another 6-ns-long run was followed to gather the statistical quantities. Finally, the ion and water density profiles perpendicular to the lower plate were obtained.

**Results and discussions**

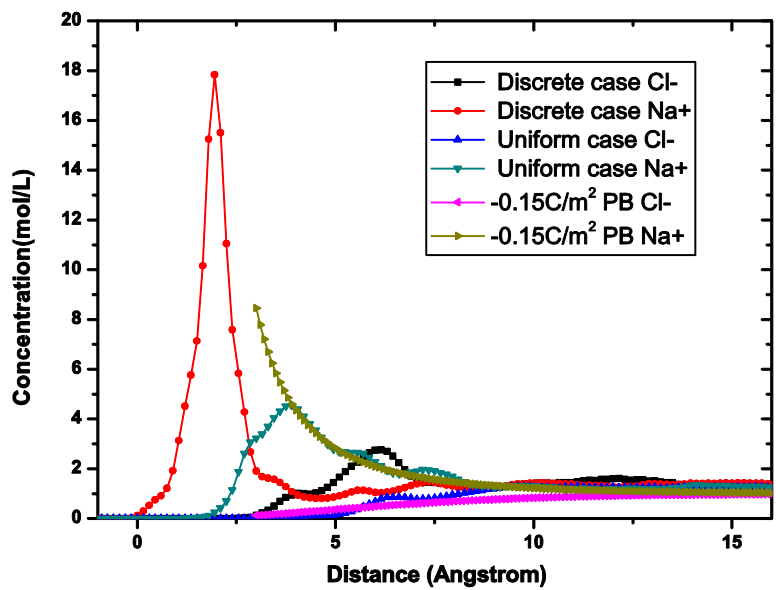

Figure 3. The ions distributions in the nanochannels with the two different charge distributions along with PB prediction.

The ion distributions in two equally charged nanochannels with different charge distributions are compared with the mean field theory as shown in Fig. 3. It can be found that at the interface with discrete surface charges the counterions $Na^+$ ions approach much nearer to the surface. The accumulating layer of $Na^+$ ions locates at 0.19 nm far away from the surface which is almost a hydration radius distance[34]. The concentration of $Na^+$ ions at the accumulation layer is as high as 17.8 mol/L and drop to the bulk concentration sharply beyond 0.3 nm. While in the uniformly charged system, the location of the main accumulating layer is 0.38 nm, and the concentration exponentially decreases to the bulk value after that. The numerical solution of PB equation is shifted to fit the ion distribution obtained in the uniformly charged nanochannel[1]. From the comparison between numerical solution and the MD data, the ion distributions near uniformly charged surface agree well with the theoretical prediction. Due to the ignorance of the ion size and the accumulating way, there is obvious discrepancy at the beginning of the PB prediction. However, the distributions of ions at discretely charged surface do not follow the PB prediction until 0.75 nm far away from the surface. This is mainly attributed to the correlation among ions and surface charges which is not considering in the mean field theory[35]. In negatively charged channels, $Cl^-$ ions act as co-ions which have the same sign as the surface. In the uniformly charged nanochannel, $Cl^-$ ions appear at 0.3 nm and its concentration increases slowly to the bulk concentration. While in discretely charged system, the concentration of $Cl^-$ ions rises fast and reaches its maximum at 0.6 nm which exceeds that of $Na^+$ ion, and decreases to the bulk concentration after that. This is because the charge inversion (see following text) caused by the overscreening of $Na^+$ ions. The potential at 0.6 nm away from the surface is positive and the effective counterions are $Cl^-$ ions in that region.

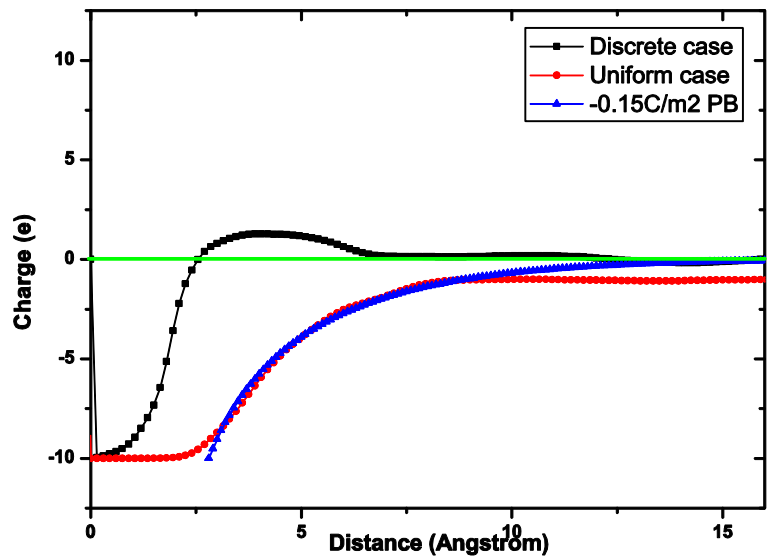

Figure 4. The integrated charge distributions perpendicular to the surfaces under two different charge distributions along with PB prediction.

Figure 4 exhibits the integrated charge distributions obtained in two MD simulations with different charge distributions and the PB equation. The integrated charge distribution perpendicular to the surface can be calculated through the equation:

$$\sigma_{IC}(i) = -N_s + \sum_{1}^{i}\left[N_{Na^+}(t) - N_{Cl^-}(t)\right]$$

Where $\sigma_{IC}(i)$ is the integrated charge accumulating to position $i$ from the surface, $N_s$ is the number of surface charge, $N_{Na^+}(t)$ and $N_{Cl^-}(t)$ are the local number of Na$^+$ and Cl$^-$ ions at position $t$. The surface charges of the simulations are set at 0 in the horizontal axis. The PB prediction is calculated from the ion concentrations used in Fig. 3. Through the comparison among the three profiles, the net charge integrated from uniformly charged surface decreases to neutrality from negative value and can be perfectly described by the PB equation. However, in this nanochannel, the electrical neutrality cannot reach which is due to the weak attraction between the surface charges and the counterions (The neutrality reached in the whole system). While, the profile near discretely charged surface is obviously different from the PB prediction. Due to the discrete surface charge which is equal to elementary charge amount, there is strong correlation among the surface charges and the counterions. Na$^+$ ions approach much nearer to the surface and the surface charges are totally screened at 0.26 nm where the Na$^+$ ions begin to appear at the uniform surface. Farther away from the surface, the effective charge becomes positive due to the overscreening of Na$^+$ ions that is the so-called charge inversion. The percent of the inverse charge amount is as high as 12.8% of the surface charges. When charge inversion occurs, the original co-ions in the channel i.e. Cl$^-$ ions act as the counterions[18].

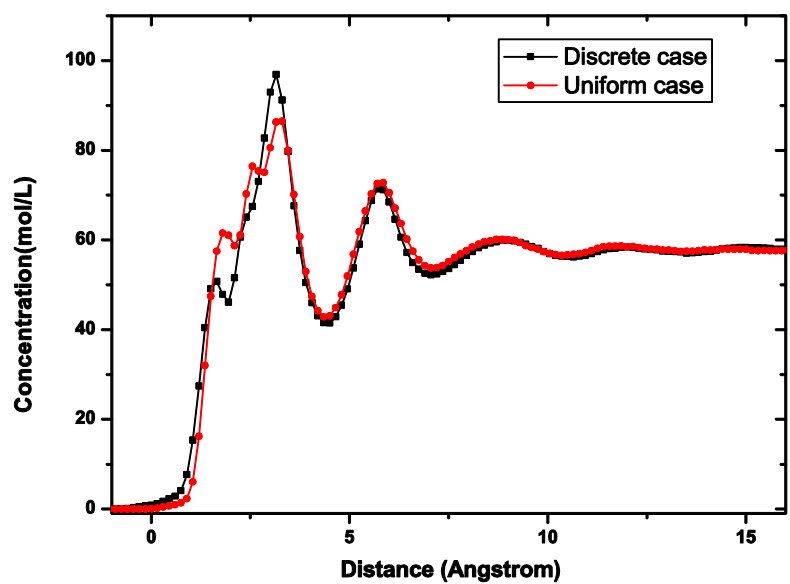

Figure 5. The water density profiles in discretely and uniformly charged nanochannels.

The water density profiles in the discretely and uniformly charged channels are plot in Fig. 5. The two profiles show similar trends except the negligible discrepancy within 0.33 nm from the surface. Within 1.0 nm from the surface, water molecules have an oscillatory distribution which reflects the phase change from the solid to liquid. This phenomenon is attributed to the interface effect[36]. The concentration of the water beyond 1.0 nm is 56 mol/L equals to the bulk concentration value. Based on the two similar water distributions, we think the surface charge distribution has negligible influence on the water density profile.

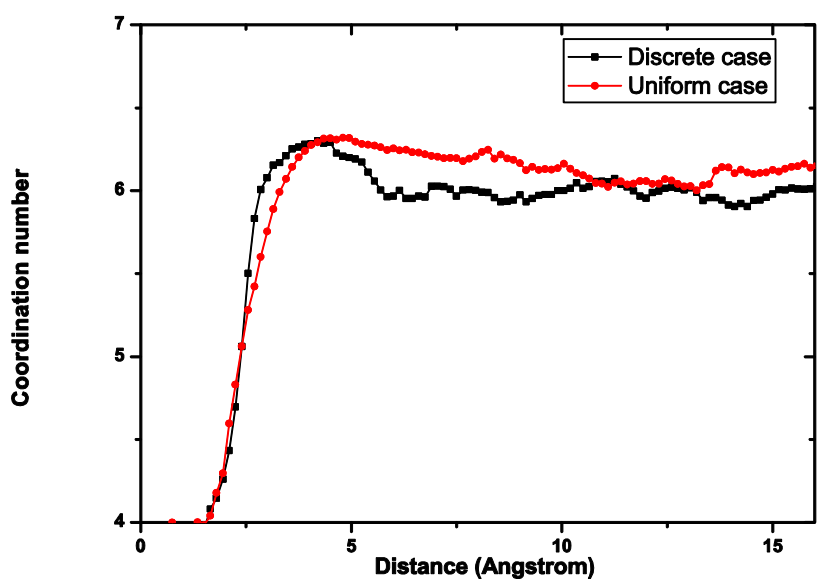

Figure 6. The distributions of coordination number per $Na^+$ ion perpendicular to the surfaces.

The hydration performance in the discretely and uniformly charged nanochannels is investigated by calculating the coordination number of ions within the first hydration shell[34]. The coordination numbers of ions locating at different distance from the surface are shown in Fig. 6. We found that the coordination number of $Na^+$ ions has a larger value near the surface and smaller one in the center. The difference of the two profiles is the decreasing way: In discrete case, the $Na^+$ ions coordination numbers drops sharply to the centre value after it reaches the maximum. While, it decreases much more slow in uniform case. The discrepancy between the two distributions is mainly because the density distribution of $Cl^-$ ions. Where $Cl^-$ ions concentration is high, the coordination number of $Na^+$ ions is low. The average coordination numbers of $Na^+$ ions are 5.2 and 6.1 in the discretely and uniformly charged nanochannels, respectively, which shows that the discrete surface charges have larger confined effect on the ions than the uniform charges.

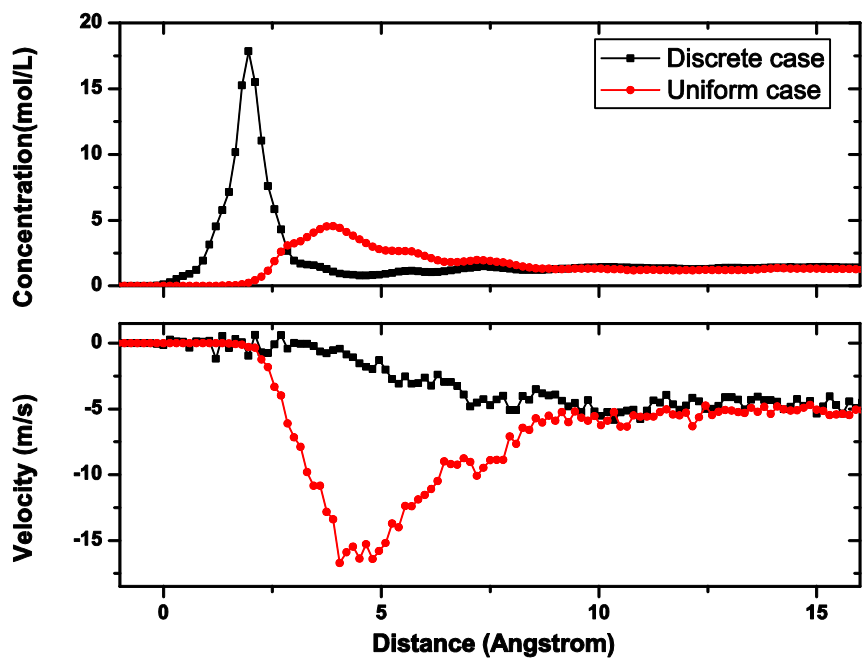

Figure 7. The concentration and velocity distributions of Na$^+$ ions in nanochannels with different surface charge distributions.

When the bulk and buffer regions on both sides of the MD model removed, an electric field with its strength 0.75 V/nm was applied across the nanochannel (see Fig.S1). The ions flowed in the specified directions. The velocity distributions in the two models were analyzed, and the results are shown in Fig. 7. It is found that Na$^+$ ions in the main accumulating layer near discretely charged surface kept stable. The ion velocity rises gradually with the distance between it and the surface increasing. In the center of the nanochannel, the Na$^+$ ions have the largest velocity. While in the uniformly charged nanochannel, the Na$^+$ ions begin to move at the location 0.2 nm from the surface and their velocity show an increase-decrease profile. When the Na$^+$ ions locate at 0.45 nm, they have the largest velocity. In the center of the channel, the velocity of ions dropped to 5.0 m/s equal to that in the discretely charged channel. The discrepancy between the two velocity distributions is mainly attributed to the effects of the surface charges and water distribution on the counterions. When the surface charge is discrete, the surface charges have strong attraction to the counterions in the vicinity of the surface (see Fig.S3). So $Na^+$ ions cannot move if they locate too near the surface. As the ions stay farther from the surface, this attractive force deceases. The velocity increases gradually. While the surface charge is uniformly distributed, each charged silicon atoms have small charge value and the interaction between surface charges and ions is very smooth as the ions translocate parallel to the surface. The velocity peak of Na+ ion is mainly because of the small hindrance of water molecules at 0.45 nm where $Na^+$ peak locates between two water layers (see Fig. S4)[37].

**Conclusion**

The key part in lab-on-a-chip technology electro-osmotic flow has close relation with the ion and water distributions at the solid-liquid interface which determines the flow character of fluid confined in nanochannels. Via MD simulations, we investigated the effect of discrete surface charge on the ion and water distributions. Results show that with the degree of the surface charge discreteness increasing, more counterions accumulate near the surface in a denser layer. Charge inversion appears due to the overscreening of $Na^+$ ions which have potential application in DNA therapy. However, the surface charge distributions have negligible effect on the water distribution but influence the hydration performance obviously. From the electro-osmotic flow in the nanochannels with different charge distributions, the discrete surface charges have large confinement to the solution flow. Uniformly charged surface can enlarge the slip length of solution and enhance the flow in the nanochannel. The simulation of aqueous solution confined in nanochannel can increase our understanding to nanofluid. And the clear physics picture of solid-liquid interface can provide some advice to the design of devices based on nanofluid.

Acknowledgements

This work was supported by the National Basic Research Program of China under Grant numbers 2011CB707601 and 2011CB707605, the National Natural Science Foundation of China under Grant number 50925519, the Fundamental Research Funds for the Central Universities, the Innovative Project for Graduate Students of Jiangsu Province under Grant number CXZZ13_0087, and the Scientific Research Foundation of Graduate School of Southeast University under Grant number YBJJ 1322. The calculations were performed on Tianhe-1A at National Supercomputing Centre in Tianjin, China.